\documentclass[11pt,twoside]{article}

\usepackage{asp2006}
\usepackage{epsf}
\usepackage{psfig}
\usepackage{lscape}

\begin{document}
\title{Possible Relic Lobes in Giant Radio Sources}   
\author{S. Godambe,$^{1,3}$ C. Konar,$^2$ D. J. Saikia,$^3$ and P. J. Wiita$^{4,5}$}  
\affil{$^1$The University of Utah, Salt Lake City, USA \\
$^2$IUCAA, Pune University Campus, Pune 411007, India \\
$^3$National Centre for Radio Astrophysics, TIFR, Pune 411007, India  \\  
$^4$Georgia State University, Atlanta, USA \\
$^5$Institute for Advanced Study, Princeton, USA}

\begin{abstract} 
We present low-frequency observations with the GMRT of
three giant radio sources (J0139+3957, J0200+4049 and J0807+7400) with relaxed diffuse
lobes which show no hotspots and no evidence of jets. The largest of these three,
J0200+4049, exhibits a depression in the centre of the western lobe, while
J0139+3957 and  J0807+7400 have been suggested earlier by Klein et al. and Lara et al.
respectively to be relic radio sources.  We estimate the spectral ages
of the lobes.  All three sources have compact radio cores.
Although the radio cores suggest that the sources are currently active, we suggest that 
the lobes in these sources could be due to an earlier cycle of activity.
\end{abstract}


\section{Introduction}   
Giant radio sources (GRSs), defined to be over $\sim$1 Mpc in size
(H$_o$=71 km s$^{-1}$ Mpc$^{-1}$, $\Omega_m$=0.27, $\Omega_{vac}$=0.73), could
provide valuable insights towards understanding the late stages of evolution
of radio sources. In the course of our study of GRSs (Ishwara-Chandra \& Saikia 1999;
Konar et al. 2004, 2008; Jamrozy et al. 2008), a number of these objects 
appeared to have diffuse lobes of emission without any prominent peaks of emission 
towards the outer edges of the lobes. Unlike Fanaroff-Riley Class I sources, 
these do not have any jets and the radio emission could be from relic lobes.
We have chosen three of these sources, namely
J0139+3957 (4C39.04), J0200+4049 (4C40.08) and J0807+7400, for a detailed investigation.
Further details are in Godambe et al. (2009).

 \begin{figure}
   \hspace{1.2cm}
   \hbox{
 \psfig{file=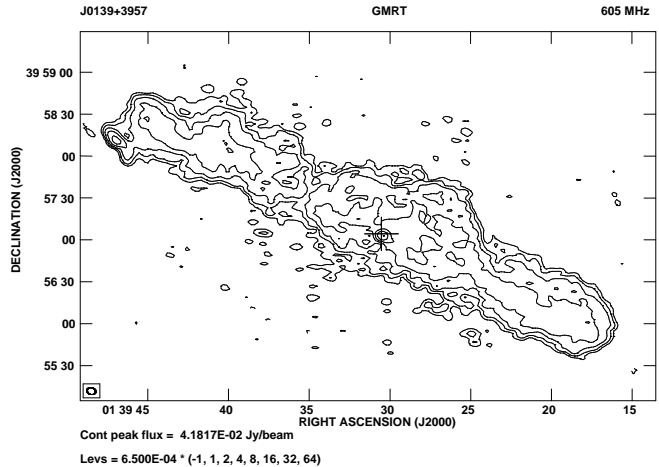,width=3.5in,angle=-90}
        }
 \caption{GMRT image of J0139+3957 at 605 MHz
 with an angular resolution of $\sim$6 arcsec. The
   $+$ sign indicates the position of the parent optical 
   galaxy. }
 \end{figure}

\section{Notes on the sources}
{\bf J0139+3957:} The large-scale structure shows the
relaxed lobes (Figure 1; Klein et al. 1995). This source is at a redshift of
0.2107 and has an overall size of 1259 kpc. Konar et al. (2004)
reported GMRT and VLA observations at 1287 and 4841 MHz respectively. 
The integrated spectrum of the source after subtracting the core flux
density yields an injection spectral index, $\alpha_{\rm inj}$$\sim$1.00 
for the Jaffe-Perola model using the {\tt SYNAGE} package (Murgia et al. 1999).
Here, the spectral index, $\alpha$ is defined as S$\propto\nu^{-\alpha}$. 
The fits to the western and eastern lobes adopting an injection spectral index of 1.0, 
gave spectral ages $\sim$12$^{+10}_{-1}$ and 5.3$^{+29}_{-2.2}$ Myr respectively 
using the classical estimates of magnetic field (Konar et al. 2008).

\noindent
{\bf J0200+4049:}
This radio galaxy is located at a redshift of 0.0827
and its largest linear size is 1414 kpc.
Our GMRT image at 333 MHz (Figure 2) shows the diffuse lobes of emission
with  evidence of a depression in the centre of the western lobe.
This clearly suggests that the lobes are no longer being fed with energy from
the nucleus.  It is also seen at 239 MHz where the data is of poorer quality,
and at 605 MHz which require more short-spacing data to produce a better
image. The flux density of the source appears to have been under-estimated in
the 74-MHz VLSS image and at 178 MHz in the 4C survey, possibly due to
the large extent of the diffuse low-brightness lobes of emission. The
values at these frequencies lie significantly below the fit to the integrated spectrum.
The injection spectral index has been estimated to be 0.73 without including
measurements which have missed a large fraction of the flux density (Figure 3). 
The spectral ages estimated for the western and eastern lobes are
$\sim$143$^{+25}_{-5}$ and 151$^{+25}_{-5}$ Myr respectively,
which is the largest amongst the three sources discussed here.

 \begin{figure}
   \hspace{1.2cm}
   \hbox{
 \psfig{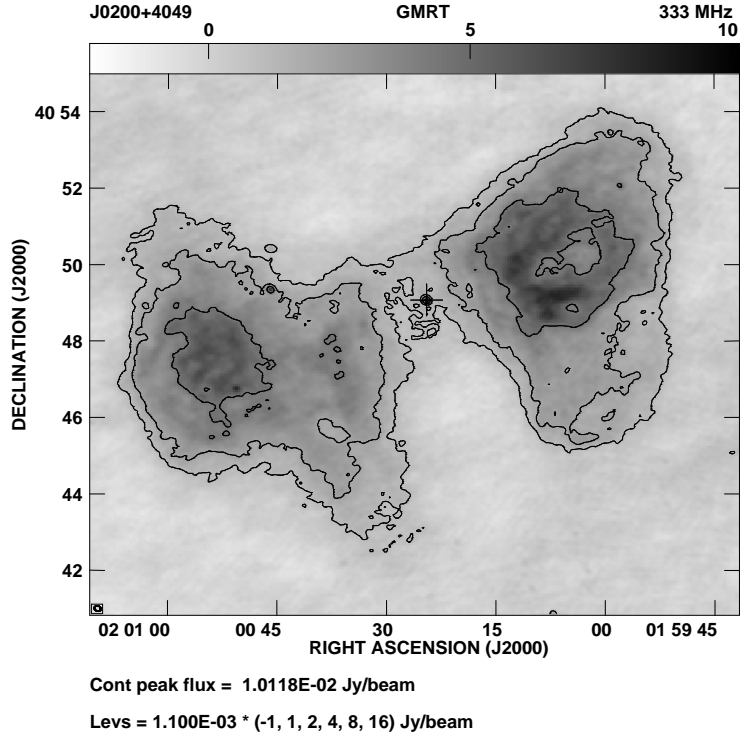}
        }
 \caption{GMRT image of J0200+4049 at 333 MHz with an angular resolution
of $\sim$9 arcsec. }
 \end{figure}

 \begin{figure}
   \hbox{
 \psfig{file=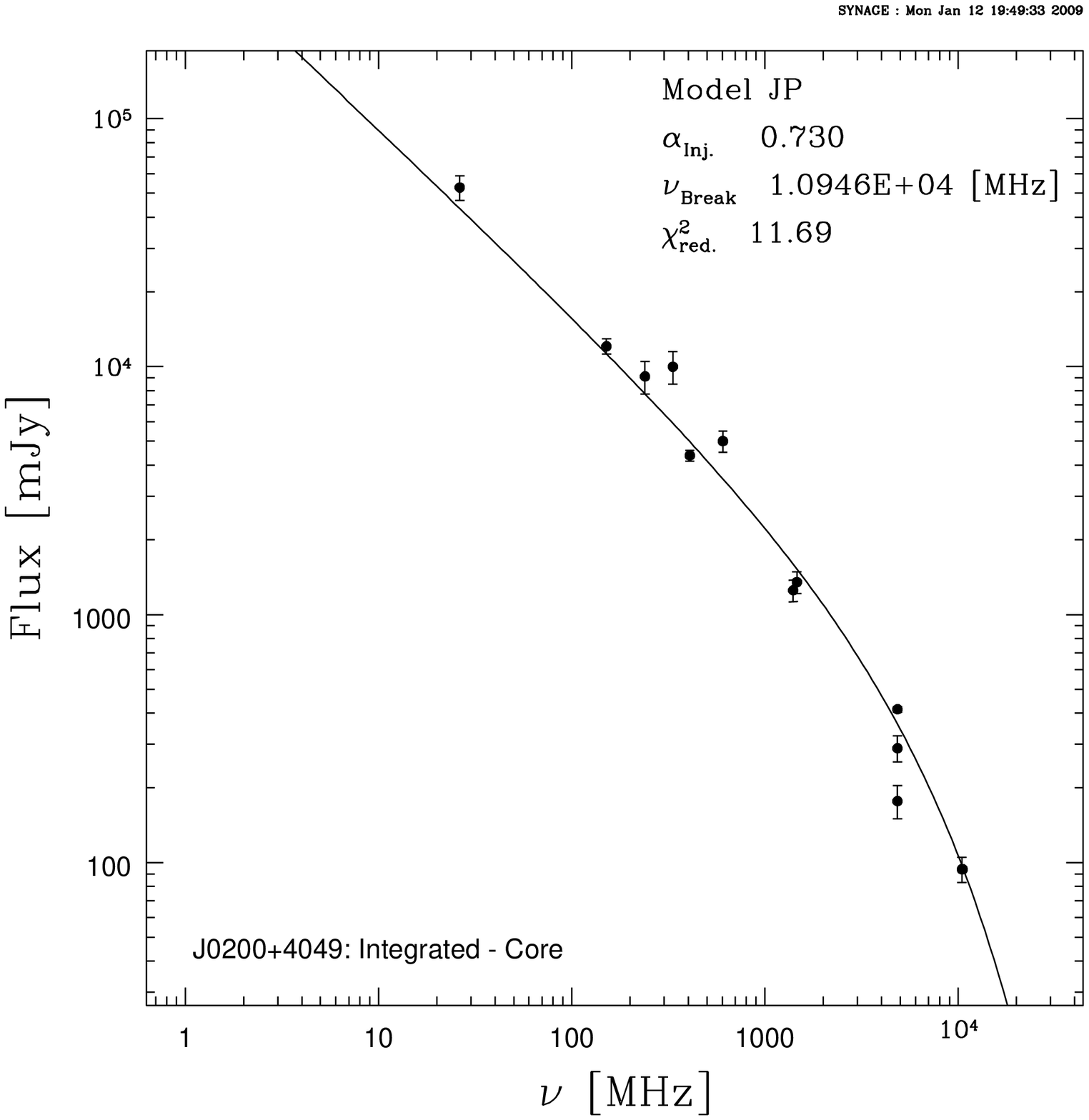,width=1.8in,angle=0}
 \psfig{file=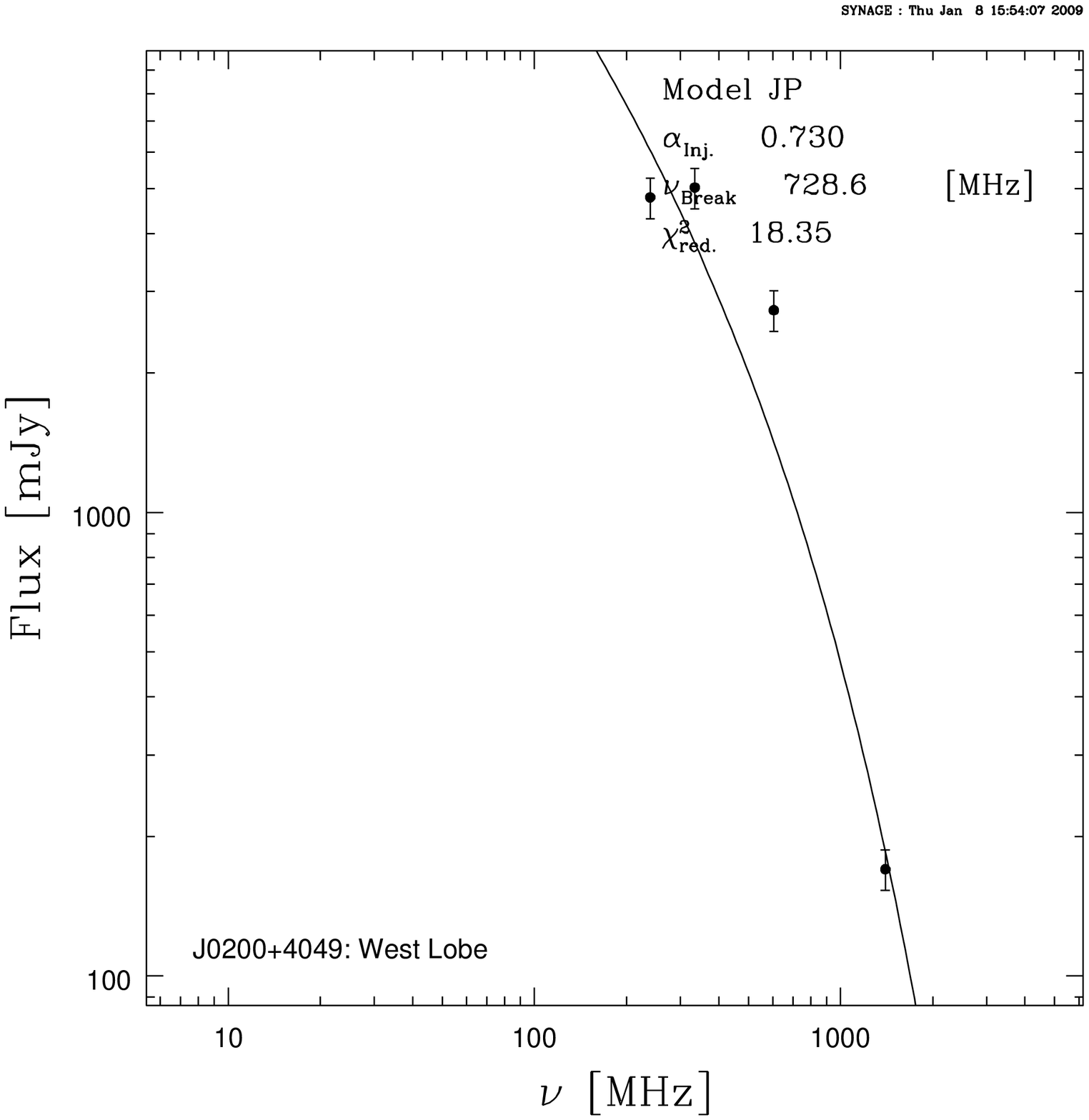,width=1.8in,angle=0}
 \psfig{file=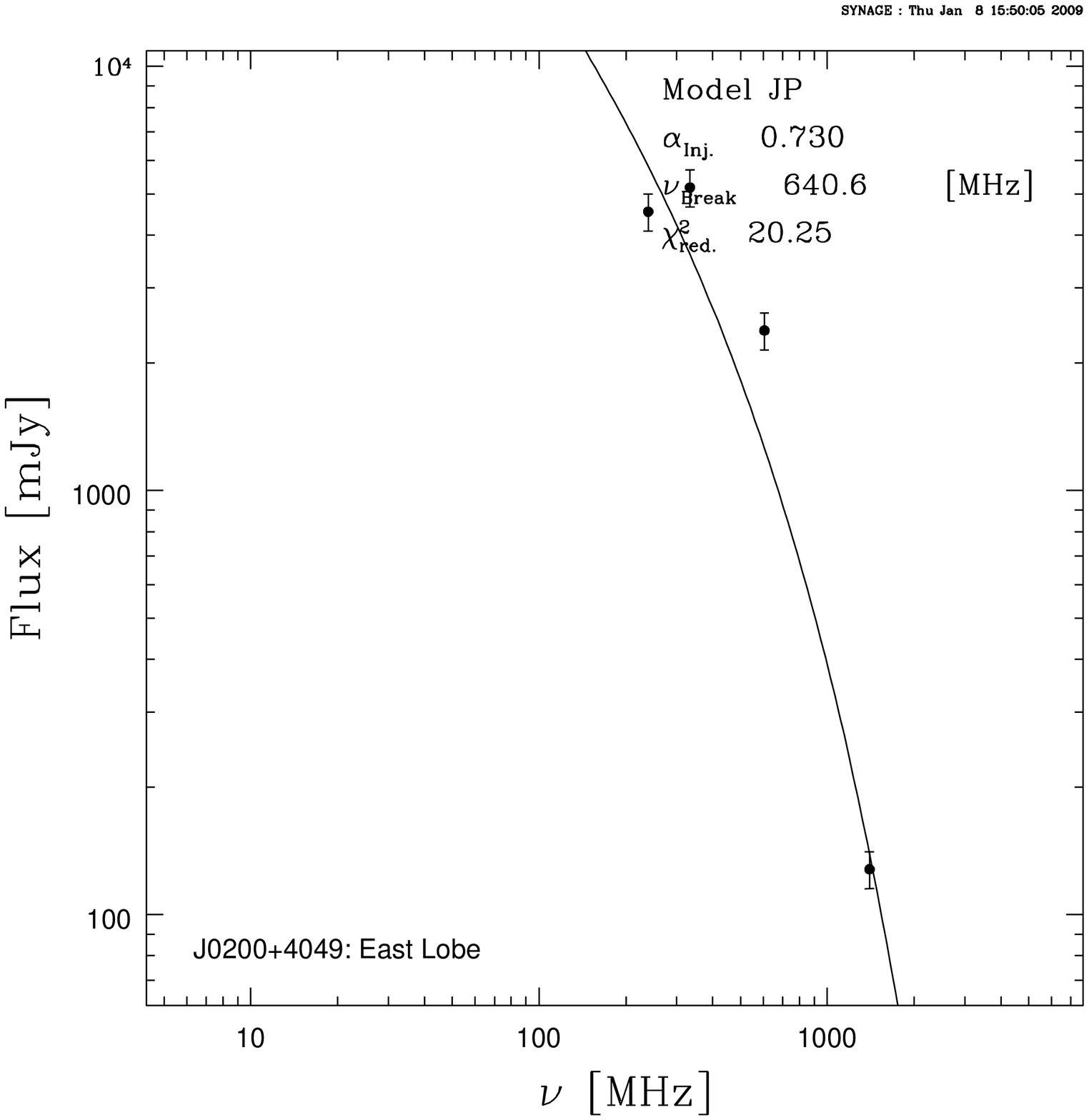,width=1.8in,angle=0}
        }
 \caption{ The fits to the spectra of the entire source after subtracting the
contribution of the core (left panel), western (middle panel) and eastern (right panel)
lobes of J0200+4049 using the {\tt SYNAGE} package (Murgia et al. 1999).
        }
 \end{figure}

\noindent
{\bf J0807+7400:}
This source is a giant low-power radio galaxy, the weakest of our three
GRSs. It is at a redshift of  0.1204  and its largest linear size is 1190 kpc.
Observations at 1.4 GHz by Lara et al. (2001)  show a compact
core component and a weak and extended halo-like emission elongated in the
east-west direction with no evidence of either jets or hotspots. At 4.9 GHz,
only the core component was visible. Lara et al. (2001) suggested that this object
could be a relic FR II radio galaxy where hotspot regions are no longer present.
Our GMRT images at low frequencies show the diffuse lobes and bridge of emission (Figure 4).
The spectral ages of the lobes, estimated for an injection spectral index of 0.73 are
$\sim$46$^{+57}_{-28}$ and 64$^{+24}_{-27}$ Myr for the western and eastern lobes respectively.

The core has a flat spectral index at high frequencies but appears to
have a steep radio spectrum with a spectral index of $\sim$0.6 at lower frequencies,
possibly due to unresolved jet/lobe structure from more recent activity.

 \begin{figure}
   \hspace{1.2cm}
   \hbox{
 \psfig{file=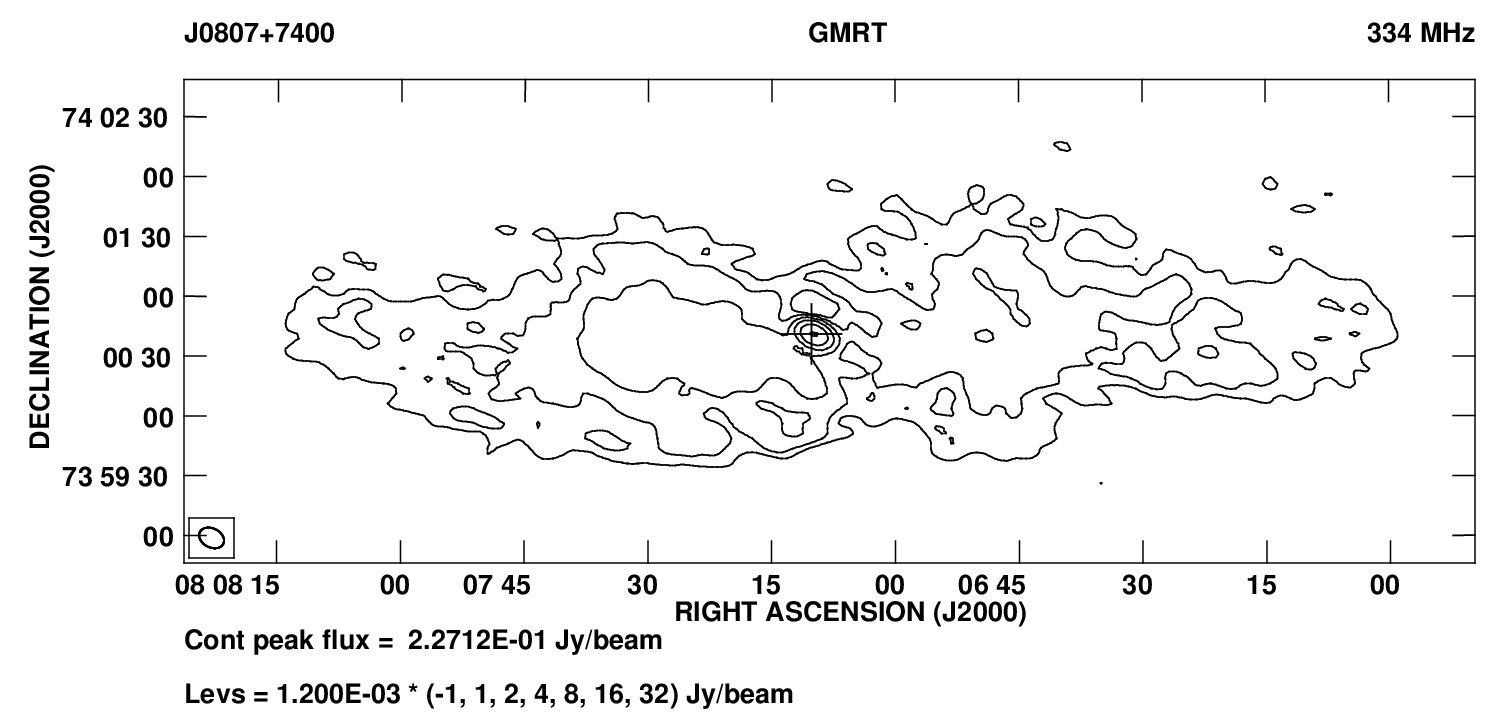,width=3.5in,angle=-90}
        }
 \caption{GMRT images of J0807+7400 at 334 MHz with an angular resolution
of $\sim$11 arcsec. }
 \end{figure}

\section{Concluding remarks}
We have explored the possibility that these sources have relic lobes by examining
the structure and spectra of all three sources
over a large frequency range. Although these three sources do not have
hotspots, their structures are not similar to the FRI sources which are
characterised by jets that expand to form the diffuse lobes of emission.
Also, the luminosities of two of the three (J0139+3957 and J0200+4049)
sources are above the dividing line for these two classes of sources.
The spectral ages are in the range of $\sim$5$\times$10$^6$ to 1.5$\times$10$^8$ yr, the upper value being
close to time scales for which the lobes are likely to remain visible if not fed with
a fresh supply of energy from the parent galaxy. However, for more reliable estimates
we need further measurements at both lower and higher frequencies.  
The structure and spectra suggest that these lobes are possibly no longer being fed,
with one of the lobes in J0200+4049 exhibiting a depression in surface brightness
towards the centre of the lobe.  The detection of cores
suggests that their nuclei are currently active. These show evidence of steep-spectrum
emission.  It would be interesting to determine the
structures of the cores from high-resolution radio observations.
Interpreting the cores as more recent activity, the time scales of episodic activity
range from $\sim$5$\times$10$^6$ to 10$^8$ yr.




\begin{thebibliography}{}
\bibitem[]{}Godambe, S., Konar, C., Saikia, D. J., \& Wiita, P.J. 2009, MNRAS, in press (arXiv:0901.3836)
\bibitem[]{}Ishwara-Chandra, C. H., \& Saikia, D. J. 1999, MNRAS, 309, 100
\bibitem[]{}Jamrozy, M., Konar, C., Machalski, J., \& Saikia, D. J. 2008, MNRAS, 385, 1286
\bibitem[]{}Klein, U., Mack, K.-H., Gregorini, L., \& Parma, P. 1995, A\&A, 303, 427
\bibitem[]{}Konar, C., Saikia, D. J., Ishwara-Chandra, C. H., \& Kulkarni, V. K. 2004, MNRAS, 355, 845
\bibitem[]{}Konar, C., Jamrozy, M., Saikia, D. J., \& Machalski, J. 2008, MNRAS, 383, 525
\bibitem[]{}Lara, L., Cotton, W. D., Feretti, L., Giovannini, G., Marcaide, J. M., M\'{a}rquez, I., \& 
            Venturi, T. 2001, A\&A, 370, 409
\bibitem[]{}Murgia, M., Fanti, C., Fanti, R., Gregorini, L., Klein, U., Mack, K.-H., \& Vigotti, M. 1999, 
            A\&A, 345, 769
\end{thebibliography}
\end{document}